\documentclass[envcountsame]{llncs} 
\everymath{\displaystyle}

\usepackage[english]{babel}
\usepackage{latexsym,amssymb,amsmath,ifthen,alltt,array,epsfig}
\usepackage{subfigure}
\usepackage[all]{xy}
\SelectTips{cm}{}

\newcommand{\premb}{\vartriangleright_\mathsf{emb}}

\newcommand{\TTT}{\textsf{T\!\raisebox{-1mm}{T}\!T}}

\newcommand{\APROVE}{\textsf{AProVE}}

\newcommand{\m}[1]{\mathsf{#1}}

\newcommand{\FF}{\mathcal{F}}

\newcommand{\NN}{\mathbb{N}}
\newcommand{\RR}{\mathcal{R}}

\newcommand{\TT}{\mathcal{T}}

\newcommand{\VV}{\mathcal{V}}

\newcommand{\seq}[2][n]{{#2_1},\dots,{#2_{#1}}}

\newcommand{\ENCODE}[3]{{\ulcorner\!#1#2#3\!\urcorner}}
\renewcommand{\ENCODE}[3]{{\!#1#2#3\!}}

\newcommand{\kbo}{\mathrm{kbo}}
\renewcommand{\l}{\langle}
\renewcommand{\r}{\rangle}
\newcommand{\limplies}{\rightarrow}
\newcommand{\liff}{\leftrightarrow}
\newcommand{\tsf}[1]{\textsf{#1}}

\newcommand{\KBO}{\mathit{KBO}}

\begin{document}

\title{Satisfying KBO Constraints\thanks{%
This research is supported by FWF (Austrian Science Fund) project
P18763. Some of the results in this paper were first announced in
\cite{ZM06}.}}

\author{Harald Zankl \and Aart Middeldorp}
\institute{
Institute of Computer Science\\
University of Innsbruck\\
Austria
}

\maketitle

\begin{abstract}
This paper presents two new approaches to prove termination of rewrite
systems with the Knuth-Bendix order efficiently. The constraints
for the weight function and for the precedence are encoded in
(pseudo-)propositional logic and the resulting formula is tested for
satisfiability. Any satisfying assignment represents a weight function
and a precedence such that the induced Knuth-Bendix order orients the
rules of the encoded rewrite system from left to right.
\end{abstract}

\section{Introduction}

This paper is concerned with proving termination of term rewrite
systems (TRSs) with the Knuth-Bendix order (KBO), a method
invented by Knuth and Bendix in \cite{KB70} well before termination
research in term rewriting became a very popular and competitive
endeavor (as witnessed by the annual termination competition).%
\footnote{\texttt{www.lri.fr/\~{}marche/termination-competition}}
We know of only two termination tools that contain an implementation
of KBO, \APROVE\ \cite{APROVE} and \TTT\ \cite{TTT}, but neither of
these tools incorporate KBO in their fully automatic mode for the
TRS category. This is
perhaps due to the fact that the algorithms known for deciding KBO
orientability (\cite{DKM90,KV03}) are not easy to implement
efficiently, despite the fact that the problem is known to be decidable
in polynomial time~\cite{KV03}. The aim of this paper is to make KBO a
more attractive choice for termination tools by presenting two simple
encodings of KBO orientability into (pseudo-)propositional logic such
that checking satisfiability of the resulting formula amounts to proving
KBO termination.

Kurihara and Kondo~\cite{KK04} were the first to encode a termination
method for term rewriting into propositional logic. They showed how
to encode orientability with respect to the lexicographic path order
as a satisfaction problem. Codish~\textsl{et al.}~\cite{CLS06} presented
a more efficient formulation for the properties of a precedence.
In \cite{CSLTG06,ZHM07} encodings of argument filterings are
presented which can be combined with propositional encodings of reduction
pairs in order to obtain logic-based implementations of the dependency
pair method. Propositional encodings of other termination methods are
described in \cite{EWZ06,FGMSTZ07,HW06}.

In Section~\ref{KBO:main} the necessary definitions for KBO are
presented. Section~\ref{ENC:kbo_main} introduces a purely propositional
encoding of KBO also describing the optimizations applied in the
implementation. In Section~\ref{ENC:pkbo_main} an alternative
encoding is given using pseudo-boolean constraints. We compare the 
power and run times of our implementations with the ones of \APROVE\ and
\TTT\ in Section~\ref{EXP:main} and show the enormous gain in efficiency.
We draw some conclusions in Section~\ref{COM:main}. One of these is
that our pseudo-boolean encoding of KBO revealed a bug in MiniSat+.
Section~\ref{SUM:main} summarizes the main contributions of this paper.

\section{Preliminaries}
\label{KBO:main}

We assume familiarity with the basics of term rewriting
(e.g.\ \cite{BN98}). In this preliminary section we recall the
definition of KBO.
A \emph{quasi-precedence} $\succsim$ (\emph{strict precedence} $\succ$)
is a quasi-order (proper order) on a signature $\FF$. Sometimes we find
it convenient to call a quasi-precedence simply precedence.
A \emph{weight function} for a signature $\FF$ is a pair $(w,w_0)$
consisting of a mapping $w\colon \FF \to \NN$ and a constant $w_0 > 0$
such that $w(c) \geqslant w_0$ for every constant $c \in \FF$.
\label{PRE:weight}
Let $\FF$ be a signature and $(w,w_0)$ a weight function for $\FF$. The
\emph{weight} of a term $t \in \TT(\FF,\VV)$ is defined as follows:
$$
w(t) = \begin{cases}
w_0 & \text{if $t$ is a variable,} \\
w(f) + \sum_{i=1}^n w(t_i) & \text{if $t = f(\seq{t})$.}
\end{cases}
$$
A weight function $(w,w_0)$ is \emph{admissible} for a quasi-precedence
$\succsim$ if $f \succsim g$ for all function symbols $g$ whenever $f$ 
is a unary function symbol with $w(f) = 0$. 
For a term $t$ $|t|$ denotes its length, i.e., the number of symbols and
$|t|_x$ ($|t|_f$) denotes how often the variable $x$ (function symbol $f$)
occurs in $t$.

\begin{definition}[{\rm\cite{KB70,DKM90,S89}}]
Let $\succsim$ be a quasi-precedence and $(w,w_0)$ a weight function.
We define the \emph{Knuth-Bendix order} $>_\kbo$ on terms inductively as
follows: $s >_\kbo t$ if $|s|_x \geqslant |t|_x$ for all variables
$x \in \VV$ and either
\begin{itemize}
\item[(a)]
$w(s) > w(t)$, or
\item[(b)]
$w(s) = w(t)$ and one of the following alternatives holds:
\begin{itemize}
\item[(1)]
$t \in \VV$, $s \in \TT(\FF^{(1)},\{ t \})$, and $s \neq t$, or
\item[(2)]
$s = f(\seq{s})$, $t = g(\seq[m]{t})$, $f \sim g$, and there exists an
$1 \leqslant i \leqslant \min \{ n, m \}$ such that $s_i >_\kbo t_i$ and
$s_j = t_j$ for all $1 \leqslant j < i$, or
\item[(3)]
$s = f(\seq{s})$, $t = g(\seq[m]{t})$, and $f \succ g$.
\end{itemize}
\end{itemize}
where $\FF^{(n)}$ denotes the set of all function symbols $f \in \FF$ of
arity $n$.
Thus in case $(b)(1)$ the term $s$ consists of a nonempty
sequence of unary function symbols applied to the variable $t$.
\end{definition}

Specializing the above definition to (the reflexive closure of) a
strict precedence, one obtains the definition of KBO in \cite{BN98},
except that we restrict weight functions to have range $\NN$ instead of
$\mathbb{R}$. According to \cite{KV03} this does not decrease the power
of the order.

\begin{lemma}
A TRS $\RR$ is terminating whenever there exist
a quasi-precedence $\succsim$ and a weight function
$(w,w_0)$ such that
$\RR \subseteq {>_\kbo}$.
\qed
\end{lemma}

\begin{example}
\label{example}
The TRS \textsf{SK\_90.2.42}\footnote{Labels in \textsf{sans-serif}
font refer to TRSs in the Termination Problems Data Base~\cite{TPDB}.}
consisting of the rules
\newcommand{\cc}{\mathop{+\!\!+}}
\begin{xalignat*}{2}
\m{flatten}(\m{nil}) &\to \m{nil} &
\m{rev}(\m{nil}) &\to \m{nil} \\
\m{flatten}(\m{unit}(x)) &\to \m{flatten}(x) &
\m{rev}(\m{unit}(x)) &\to \m{unit}(x) \\
\m{flatten}(x \cc y) &\to \m{flatten}(x) \cc \m{flatten}(y) &
\m{rev}(x \cc y) &\to \m{rev}(y) \cc \m{rev}(x) \\
\m{flatten}(\m{unit}(x) \cc y) &\to \m{flatten}(x) \cc \m{flatten}(y) &
\m{rev}(\m{rev}(x)) &\to x \\
\m{flatten}(\m{flatten}(x)) &\to \m{flatten}(x) &
(x \cc y) \cc z &\to x \cc (y \cc z) \\
x \cc \m{nil} &\to x &
\m{nil} \cc y &\to y
\end{xalignat*}
is 
KBO terminating. The weight function $(w,w_0)$ with
$w(\m{flatten}) = w(\m{rev}) = w(\cc) = 0$ and
$w(\m{unit}) = w(\m{nil}) = w_0 = 1$ together with the
quasi-precedence
$\m{flatten} \sim \m{rev} \succ \m{unit} \succ \m{\cc} \succ \m{nil}$
ensures that $l >_\kbo r$ for all rules $l \to r$. The use of a
quasi-precedence is essential here; the rules
$\m{flatten}(x \cc y) \to \m{flatten}(x) \cc \m{flatten}(y)$ and
$\m{rev}(x \cc y) \to \m{rev}(y) \cc \m{rev}(x)$ demand
$w(\m{flatten}) = w(\m{rev}) = 0$ but KBO with strict precedence
does not allow different unary functions to have weight zero.
\end{example}

One can imagine a more general definition of KBO. For instance, in 
case (b)(2) we could demand that
$s_j \sim_\kbo t_j$ for all $1 \leqslant j < i$ where $s \sim_\kbo t$
if and only if $s \sim t$ and $w(s) = w(t)$. Here $s \sim t$ denotes 
syntactic equality with respect to equivalent function symbols of the
same arity.
Another obvious extension would be to compare the arguments
according to an arbitrary permutation or as multisets.
To keep the discussion and implementation simple, we do not consider
such refinements in the sequel.

\section{A Pure SAT Encoding of KBO}
\label{ENC:kbo_main}

\newcommand{\xor}{\oplus}
In order to give a propositional encoding of KBO termination,
we must take care of representing a precedence and a weight
function. For the former we introduce two sets of new variables
\(
X = \{ X_{fg} \mid \text{$f, g \in \FF$ with $f \neq g$} \}
\)
and
\(
Y = \{ Y_{fg} \mid \text{$f, g \in \FF$ with $f \neq g$} \}
\)
depending on the underlying signature $\FF$ (\cite{KK04,Z06}).
The intended semantics of these variables is that an assignment which
satisfies a variable $X_{fg}$ corresponds to a precedence with
$f \succ g$ and similarly $Y_{fg}$ suggests $f \sim g$. 
When dealing with strict precedences it is safe to assign all 
$Y_{fg}$ variables to false. 
For the weight function, symbols are considered in binary
representation and the operations $>$, $=$, $\geqslant$, and $+$ must
be redefined accordingly. The propositional encodings of $>$ and $=$
given below are similar to the ones in \cite{CLS06}.
To save parentheses we employ the binding hierarchy for the
connectives where $+$ binds strongest, followed by the relation
symbols $>$, $=$, and $\geqslant$. The logical connectives $\lor$ and
$\land$ are next in the hierarchy and $\limplies$ and $\liff$ bind weakest.

We fix the number $k$ of bits that is available for representing
natural numbers in binary. Let $a < 2^k$. We denote by
$\mathbf{a} = \l a_k, \dots, a_1 \r$ the binary representation of $a$
where $a_k$ is the most significant bit.

\begin{definition}
For natural numbers given in binary representation, the operations $>$,
$=$, and $\geqslant$ are defined as follows (for all
$1 \leqslant j \leqslant k$):
\begin{align*}
\mathbf{f} >_j \mathbf{g} &~=~
\begin{cases}
f_1 \land \lnot g_1 & \text{if $j = 1$} \\
(f_j \land \lnot g_j) \lor \bigl((f_j \leftrightarrow g_j) \land
\mathbf{f} >_{j-1} \mathbf{g}\bigr) & \text{if $j > 1$}
\end{cases} \\
\mathbf{f} > \mathbf{g} &~=~
\mathbf{f} >_k \mathbf{g} \\
\mathbf{f} = \mathbf{g} &~=~
\bigwedge_{i=1}^k (f_i \leftrightarrow g_i) \\
\mathbf{f} \geqslant \mathbf{g} &~=~
\mathbf{f} > \mathbf{g} \lor \mathbf{f} = \mathbf{g}
\end{align*}
\end{definition}

Next we define a formula which is satisfiable if and only if the encoded
weight function is admissible for the encoded precedence.

\begin{definition}
For a weight function $(w,w_0)$, let $\m{ADM}\text{-}\m{SAT}(w,w_0)$ be
the formula
\[
\mathbf{w_0} > \mathbf{0} ~ \land
\bigwedge_{c\in \FF^{(0)}} \mathbf{c} \geqslant \mathbf{w_0}
~ \land \bigwedge_{f \in \FF^{(1)}}
\big(\mathbf{f} = \mathbf{0} ~ \limplies
\bigwedge_{\makebox[1cm]{$\scriptstyle g \in \FF,\,f \neq g$}}
(X_{fg} \lor Y_{fg}) \big)
\]
\end{definition}

For addition we use pairs. The first component represents the bit
representation and the second component is a propositional formula which
encodes the constraints for each digit.

\begin{definition}
We define $(\mathbf{f}, \varphi) + (\mathbf{g}, \psi)$
as $(\mathbf{s}, \varphi \land \psi \land \gamma \land \sigma)$ with
\[
\gamma ~=~ \lnot c_k \land \lnot c_0 \land \bigwedge_{i=1}^{k}
\big(c_i \liff ((f_i \land g_i) \lor (f_i \land c_{i-1}) \lor
(g_i \land c_{i-1}))\big)
\]
and
\[
\sigma ~=~ \bigwedge_{i=1}^k \big( s_i \liff
(f_i \xor g_i \xor c_{i-1})\big)
\]
where $c_i$ $(0 \leqslant i \leqslant k)$ and $s_i$
$(1 \leqslant i \leqslant k)$ are fresh variables that represent the
carry and the sum of the addition and
$\xor$ denotes exclusive or. The condition $\lnot c_k$ prevents a possible
overflow.
\end{definition}

Note that although theoretically not necessary, it is a good idea to
introduce new variables for the sum. The reason is that in
consecutive additions each bit $f_i$ and $g_i$ is duplicated (twice for
the carry and once for the sum) and consequently using fresh variables
for the sum prevents an exponential blowup of the resulting formula.

\begin{definition}
We define $(\mathbf{f},\varphi) > (\mathbf{g}, \psi)$ as
$\mathbf{f} > \mathbf{g} \land \varphi \land \psi$. 
Equality is defined similarly where 
$(\mathbf{f},\varphi) = (\mathbf{g}, \psi)$ is 
$\mathbf{f} = \mathbf{g} \land \varphi \land \psi$.
\end{definition}

In the next definition we show how the weight of terms is computed
propositionally.

\begin{definition}
Let $t$ be a term and $(w,w_0)$ a weight function. The weight of a term
is encoded as follows:
\[
W_t ~=~ \begin{cases}
(\mathbf{w_0}, \top) & \text{if $t \in \VV$,} \\
(\mathbf{f}, \top) + \sum_{i=1}^{n} W_{t_i} & \text{if $t = f(\seq{t})$.}
\end{cases}
\]
\end{definition}

We are now ready to define a propositional formula that reflects the
definition of $>_\kbo$.

\begin{definition}
Let $s$ and $t$ be terms. We define the formula $\m{SAT}(s >_\kbo t)$
as follows. If $s \in \VV$ or $s = t$ or $|s|_x < |t|_x$ for some 
$x \in \VV$ then
$\m{SAT}(s >_\kbo t) = \bot$. Otherwise
\[
\m{SAT}(s >_\kbo t) ~=~ 
W_s > W_t \lor
\bigl(W_s = W_t \land \m{SAT}(s >_\kbo' t)\bigr)
\]
with
\[
\m{SAT}(s >_\kbo' t) = 
\begin{cases}
\top & \text{if $t \in \VV$, $s \in \TT(\FF^{(1)},\{t\})$, and $s \neq t$}
\\
\m{SAT}(s_i >_\kbo t_i) &
\text{if $s = f(\seq{s}), t = f(\seq{t})$} \\
X_{fg} \lor \bigl(Y_{fg} \land
\m{SAT}\makebox[0mm][l]{$(s_i >_\kbo t_i)\bigr)$} \\
& \hspace{-15mm}
\text{if $s = f(\seq{s})$, $t = g(\seq[m]{t})$, and $f \neq g$}
\end{cases}
\]
where in the second (third) clause $i$ denotes the least
$1 \leqslant j \leqslant n$ ($\,\min \{ n, m \}$) with
$s_j \neq t_j$.
\end{definition}

\subsection{Encoding the Precedence in SAT}
\label{ENC:symb}

To ensure the properties of a precedence we follow the approach
of Codish~\textsl{et al.}~\cite{CLS06} who propose to interpret
function symbols as natural numbers. The greater than or equal to
relation
then ensures that the function symbols are quasi-ordered. Let
$|\FF| = n$. We are looking for a mapping
$m\colon \FF \to \{ 1, \dots, n \}$ such that for every propositional
variable $X_{fg} \in X$ we have $m(f) > m(g)$ and for $Y_{fg} \in Y$
we get $m(f) = m(g)$. To uniquely encode one of the $n$ function 
symbols, $l := \lceil log_2(n) \rceil$ fresh
propositional variables are needed. The $l$-bit representation of $f$
is $\langle f'_l, \dots, f'_1 \rangle$ with $f'_l$ the most significant
bit.

\begin{definition}
For all $1 \leqslant j \leqslant l$
\begin{align*}
||X_{fg}||_j &~=~
\begin{cases}
f'_1 \land \lnot g'_1 & \text{if $j = 1$} \\
(f'_j \land \lnot g'_j) \lor \bigl((f'_j \leftrightarrow g'_j) \land
||X_{fg}||_{j-1}\bigr) & \text{if $j > 1$}
\end{cases} \\[2ex]
||Y_{fg}||_l &~=~ \bigwedge_{j=1}^l (f'_j \liff g'_j)
\end{align*}
\end{definition}

Note that the variables $f'_i \ (1 \leqslant i \leqslant l)$ are 
different from $f_i \ (1 \leqslant i \leqslant k)$ which are used
to represent weights.

\begin{definition}
Let $\RR$ be a TRS. The formula $\m{KBO}$-$\m{SAT}(\RR)$ is defined as
\[
\m{ADM}\text{-}\m{SAT}(w,w_0) ~ \land \bigwedge_{l \to r \in \RR}
\m{SAT}(l >_\kbo r) ~ \land
\bigwedge_{z \in X \cup Y} (z \leftrightarrow ||z||_l)
\]
\end{definition}

\begin{theorem}
\label{kbothm1}
A TRS $\RR$ is terminating whenever the propositional formula
$\m{KBO}$-$\m{SAT}(\RR)$ is satisfiable.
\qed
\end{theorem}

The reverse does not hold (Example~\ref{unbounded bits}).

\subsection{Optimizations}
\label{OPT:main}

This section deals with logical simplifications concerning propositional
formulas as well as optimizations which are specific for the generation
of the constraint formula which encodes KBO termination of the given
instance.

\subsubsection{Logical Optimizations}
\label{OPT:log}

Since the constraint formula contains many occurrences of $\top$ and
$\bot$ logical equivalences simplifying such formulas are employed.


SAT solvers typically expect their input in conjunctive normal
form (CNF) but for the majority of the TRSs the constraint formula
$\m{KBO}$-$\m{SAT}(\RR)$ is too large for the standard translation.
The problem is that the resulting CNF may be exponentially 
larger than the input formula because when distributing $\lor$ over
$\land$ subformulas get duplicated.
In \cite{T68} Tseitin proposed a transformation which is linear in the
size of the input formula. The price for linearity
is paid with introducing new variables. As a consequence,
Tseitin's transformation does not produce an equivalent formula,
but it does preserve and reflect satisfiability.

\subsubsection{Optimizations Concerning the Encoding}
\label{OPT:enc}

Before discussing the implemented optimizations in detail it is worth
mentioning the bottleneck of the whole procedure. As addressed in the
previous section, SAT solvers expect their input in CNF. It turned out
that the generation of all non-atomic subformulas, which are needed for
the translation, constitutes the main bottleneck. So every change in
the implementation which reduces the size of the
constraint formula will result in an additional speedup.
All improvements discussed in the sequel could reduce the execution
time at least a bit. Whenever they are essential it is explicitly
stated.

Since $>_\kbo$ is a simplification order it contains the embedding
relation. We make use of that fact by only computing the constraint
formula $\ENCODE{s}{>_\kbo}{t\,}$ if the test $s \premb t$ is false.
Most of the other optimizations deal with representing or computing
the weight function. When computing the constraints for the 
weights in a rule $l \to r$, removing function symbols and variables 
that occur both in $l$ and in $r$ is highly recommended or even 
necessary for an efficient implementation. The benefit can be 
seen in the example below. Note that propositional addition is somehow 
expensive as new variables have to be added for representing the carry 
and the sum in addition to a formula which encodes the constraints for
each digit.

\begin{example}
Consider the TRS consisting of the single rule
$\m{f}(y,\m{g}(x),x) \to \m{f}(y,x,\m{g(g}(x)))$.
Without the optimization the constraints for the weights would amount to
\begin{gather*}
(\mathbf{f},\top) + (\mathbf{w_0},\top) + (\mathbf{g},\top)
+ (\mathbf{w_0},\top) + (\mathbf{w_0},\top) \\ \geqslant \\
\qquad\qquad (\mathbf{f},\top) + (\mathbf{w_0},\top) + (\mathbf{w_0},\top)
+ (\mathbf{g},\top) + (\mathbf{g},\top) + (\mathbf{w_0},\top)
\end{gather*}
whereas employing the optimization produces the more or less trivial
constraint $(\mathbf{0},\top) \geqslant (\mathbf{g},\top)$.
\end{example}

By using a cache for propositional addition we can test if we already
computed the sum of the weights of two function symbols 
$f$ and $g$. That reduces the number
of newly introduced variables and sometimes we can omit the constraint
formula for addition. This is clarified in the following example.

\begin{example}
Consider the TRS consisting of the rules $\m{f(a)} \to \m{b}$ and
$\m{f(a)} \to \m{c}$.
The constraints for the first rule amount to the following formula
where $\mathbf{\underline{fa}}$ corresponds to the new variables
which are required for the sum when adding $\m{f}$ and $\m{a}$ and the 
propositional formula $\varphi$ represents the constraints which
are put on each digit of $\mathbf{\underline{fa}}$:
\begin{align*}
\m{SAT(f(a)} >_\kbo \m{b}) &~=~ 
W_{\m{f(a)}} > W_{\m{b}} \lor
\bigl(W_{\m{f(a)}} = W_{\m{b}} \land X_{\m{fb}}\bigr) \\
&~=~ (\mathbf{f},\top) + (\mathbf{a},\top) > (\mathbf{b},\top)
\lor \bigl((\mathbf{f},\top) + (\mathbf{a},\top) = (\mathbf{b},\top) \land
X_{\m{fb}}\bigr) \\
&~=~ (\,\mathbf{\underline{fa}},\varphi) > (\mathbf{b},\top)
\lor \bigl((\,\mathbf{\underline{fa}},\varphi) = (\mathbf{b},\top)
\land X_{\m{fb}}\bigr) \\
&~=~ (\,\mathbf{\underline{fa}} > \mathbf{b} \land \varphi) \lor
(\,\mathbf{\underline{fa}} = \mathbf{b} \land \varphi \land
X_{\m{fb}})
\end{align*}
We get a similar formula for the second rule and the conjunction of
both amounts to
\[
\bigl((\mathbf{\,\underline{fa}} > \mathbf{b} \land \varphi) \lor 
(\mathbf{\,\underline{fa}} = \mathbf{b} \land \varphi \land
X_{\m{fb}})\bigr) \land 
\bigl((\,\mathbf{\underline{fa}} > \mathbf{c} \land \varphi) \lor 
(\,\mathbf{\underline{fa}} = \mathbf{c} \land \varphi \land
X_{\m{fc}})\bigr)
\]
Using commutativity and distributivity we could obtain the equivalent
formula
\[
\bigl(\,\mathbf{\underline{fa}} > \mathbf{b} \lor 
(\,\mathbf{\underline{fa}} = \mathbf{b} \land X_{\m{fb}})\bigr)
\land 
\bigl(\,\mathbf{\underline{fa}} > \mathbf{c} \lor 
(\,\mathbf{\underline{fa}} = \mathbf{c} \land X_{\m{fc}})\bigr)
\land \varphi
\]
which gives rise to fewer subformulas. Note that this simplification
can easily be implemented using the information of the cache for addition.
\end{example}


\section{A Pseudo-Boolean Encoding of KBO}
\label{ENC:pkbo_main}

A \emph{pseudo-boolean constraint} (PBC) is of the form
\[
\big( \sum_{i=1}^n a_i * x_i\,\big) \circ m
\]
where $\seq{a}, m$ are fixed integers, $\seq{x}$ boolean variables that
range over $\{ 0, 1 \}$, and
$\circ \in \{ {\geqslant}, {=}, {\leqslant} \}$. We separate PBCs that
are written on a single line by semicolons.
A sequence of PBCs is satisfiable if there exists an assignment which 
satisfies every PBC in the sequence. Since 2005 pseudo-boolean evaluation
\cite{PBE} is a track of the international SAT
competition.\footnote{\texttt{http://sat07.ecs.soton.ac.uk}}
In the sequel we show how to encode KBO using PBCs.

\begin{definition}
\label{PSAT:adm}
For a weight function $(w,w_0)$ let $\m{ADM}$-$\m{PBC}(w,w_0)$ be the
collection of PBCs
\begin{itemize}
\item
$\overline{w}_0 \geqslant 1$
\item
$\overline{w}(c) - \overline{w}_0 \geqslant 0$ for all $c \in \FF^{(0)}$
\item
$(n-1)*\overline{w}(f) + \sum_{f \neq g} (X_{fg} + Y_{fg}) \geqslant (n-1)$
for all $f \in \FF^{(1)}$
\end{itemize}
where $n = |\FF|$, $\overline{w}(f) = 2^{k-1}*f_k + \dots + 2^0*f_1$
denotes the weight of $f$ in $\NN$ using $k$ bits,
and $\overline{w}_0$ denotes the value of $\mathbf{w_0}$.
\end{definition}

In the definition above the first two PBCs express that $w_0$ is
strictly larger than zero and that every unary function symbol
has weight at least $w_0$. 
Whenever the considered function symbol $f$ has weight larger than
zero the third constraint is trivially satisfied. In the case that the
unary function symbol $f$ has weight zero the constraints on the
precedence add up to $n-1$ if and only if $f$ is a maximal element.
Note that $X_{fg}$ and $Y_{fg}$ are mutual exclusive
(which is ensured when encoding the constraints on a quasi-precedence,
cf.\ Definition~\ref{DEF:pqprec}).

For the encoding of $s >_\kbo t$ and $s >_\kbo' t$ auxiliary propositional
variables $\KBO_{s,t}$ and $\KBO'_{s,t}$ are introduced. The intended
meaning is that if $s >_\kbo t$ ($s >_\kbo' t$) then $\KBO_{s,t}$
($\KBO'_{s,t}$) evaluates to true under a satisfying assignment. The
general idea of the
encoding is very similar to the pure SAT case. As we do not know
anything about weights and the precedence at the time of encoding we 
have to consider the cases $w(s) > w(t)$ and $w(s) = w(t)$ at
the same time. That is why $\KBO'_{s,t}$ and the recursive call to 
$\m{PBC}(s >_\kbo' t)$ must be considered in any case. 

The weight $w(t)$ of a term $t$ is defined similarly as in 
Section~\ref{PRE:weight} with the only difference that
the weight $w(f)$ of the function symbol $f \in \FF$ is represented
in $k$ bits as described in Definition~\ref{PSAT:adm}.

\begin{definition}
Let $s$ and $t$ be terms. The encoding
of $\m{PBC}(s >_\kbo t)$ amounts to $\KBO_{s,t} = 0$ if $s \in \VV$ 
or $s=t$ or $|s|_x < |t|_x$ for some $x \in \VV$. In
all other cases $\m{PBC}(s >_\kbo t)$ is 
\[
-(m + 1)*\KBO_{s,t} + w(s) - w(t) + \KBO'_{s,t} \geqslant -m; ~
\m{PBC}(s >_\kbo' t)
\]
where $m = 2^k*|t|$. Here $\m{PBC}(s >_\kbo' t)$ is the empty constraint
when $t \in \VV$, $s \in \TT(\FF^{(1)},\{t\})$, and $s \neq t$. In the
remaining case $s = f(\seq{s})$, $t = g(\seq[m]{t})$, and
$\m{PBC}(s >_\kbo' t)$ is the combination of $\m{PBC}(s_i >_\kbo t_i)$
and
\[
\begin{cases}
-\KBO'_{s,t} + \KBO_{s_i,t_i} \geqslant 0
& \text{if $f = g$} \\
-2*\KBO'_{s,t} + 2*X_{fg} + Y_{fg} + \KBO_{s_i,t_i} \geqslant 0
& \text{if $f \neq g$ }
\end{cases}
\]
where $i$ denotes the least $1 \leqslant j \leqslant \min \{ n, m \}$
with $s_i \neq t_i$.
\end{definition}

Since the encoding of $\m{PBC}(s >_\kbo t)$ is explained
in the example below here we just explain the intended semantics
of $\m{PBC}(s >_\kbo' t)$. In the first case where $t$ is a
variable there are no constraints on the weights and the precedence
which means that the empty constraint is returned. In the case where
$s$ and $t$ have identical root symbols it is demanded that
whenever $\KBO'_{s,t}$ holds then also $\KBO_{s_i,t_i}$ must be satisfied
before going into the recursion. In the last case $s$ and $t$ have
different root symbols 
and the PBC expresses that whenever $\KBO'_{s,t}$ is satisfied then
either $f > g$ or both $f \sim g$ and $\KBO_{s_i,t_i}$ must hold. 

To get familiar with the encoding and to see why the definitions are
a bit tricky consider the example below. For reasons of readability
symbols occurring both in $s$ and in $t$ are removed immediately. This
entails that the multiplication factor $m$ should be
lowered to 
\[
m = \sum_{x \in \FF \cup \VV} max \{0, 2^k*(|t|_x -|s|_x)\},
\]
which again is a lower bound of the left-hand side of the constraint if 
$\KBO_{s,t}$ is false.

\begin{example}
Consider the TRS consisting of the rule
\[
s = \m{f(g}(x),\m{g(g}(x))) \to  \m{f(g(g}(x)),x) = t
\]
The PB encoding $\m{PBC}(s >_\kbo t)$ then looks as follows: 
\begin{align}
- \KBO_{s,t} + w(g) + \KBO'_{s,t} &\geqslant 0 \\
-\KBO'_{s,t} + \KBO_{\m{g}(x),\m{g(g}(x))} &\geqslant 0 \\
- (2^k+1)*\KBO_{\m{g}(x),\m{g(g}(x))} - w(g) +
\KBO'_{\m{g}(x),\m{g(g}(x))} &\geqslant -2^k \\
\KBO'_{\m{g}(x),\m{g(g}(x))} + \KBO_{x,\m{g}(x)} &\geqslant 0 \\
\KBO_{x,\m{g}(x)} &= 0
\end{align}
Constraint (1) states that if $s >_\kbo t$ then
either $w(g) > 0$ or $s >_\kbo' t$. Clearly the
attentive reader would assign $w(g) = 1$ and termination
of the TRS is shown. The encoding however is not so smart and
performs the full recursive translation to PB. In (3) it is
not possible to satisfy $s_1 = \m{g}(x) >_\kbo \m{g}(\m{g}(x)) = t_1$ 
since the former is embedded in the latter. Nevertheless the
constraint (3) must remain satisfiable because the TRS is
KBO terminating. The trick is to introduce a hidden case 
distinction. The multiplication factor in front of the 
$\KBO_{s_1,t_1}$ variable does that job. Whenever $s_1 >_\kbo t_1$ is
needed then $\KBO_{s_1,t_1}$ must evaluate to true. Then implicitly
the constraint demands that $w(s_1) > w(t_1)$ or $w(s_1) = w(t_1)$
and $s_1 >_\kbo' t_1$ which reflects the definition of KBO. 
If $s_1 >_\kbo t_1$ need not be satisfied (e.g., because already
$s >_\kbo t$ in (1)) then the constraint holds in any case since the
left hand side in (3) never becomes smaller than $-2^k$.
\end{example}

\subsection{Encoding the Precedence in PBCs}

To encode a precedence in PB we again interpret function symbols in
$\NN$. For this approach an additional set of propositional variables
\(
Z = \{Z_{fg} \mid f, g \in \FF \text{ with } f \neq g\}
\)
is used. The intended semantics is that $Z_{fg}$
evaluates to true whenever $g \succ f$ or $f$ and $g$ are incomparable. 
Just note that the $Z_{fg}$ variables are not necessary
as far as termination proving power is considered but they
are essential to encode partial precedences which are 
sometimes handy (as explained in Section~\ref{COM:main}).

\begin{definition}
\label{DEF:pqprec}
For a signature $\FF$ we define $\m{PREC}$-$\m{PBC}(\FF)$ using the PBCs
below. Let $l = \lceil log_2(|\FF|) \rceil$. For all $f, g \in \FF$ with 
$f \neq g$
\begin{align*}
2*X_{fg} + Y_{fg} + Y_{gf} + 2*Z_{fg} = 2\\
-X_{fg} + 2^l*Y_{fg} + 2^l*Z_{fg} + i(f) - i(g) \geqslant 0\\
2^l*X_{fg} + Y_{fg} + 2^l*Z_{fg} + i(f) - i(g) \geqslant 1
\end{align*}
where $i(f) = 2^{l-1}*f'_l + \dots + 2^0*f'_1$ denotes the interpretation
of $f$ in $\NN$ using $l$ bits.
\end{definition}

The above definition expresses all requirements of a quasi-precedence.
The symmetry of $\sim$ and the mutual 
exclusion of the $X$, $Y$, and $Z$ variables is mimicked by the first
constraint. The second constraint encodes the conditions that
are put on the $X$ variables. Whenever a system needs $f > g$ in
the precedence to be terminating then $X_{fg}$ must evaluate
to true and (because they are mutually exclusive) $Y_{fg}$ and $Z_{fg}$ to
false. Hence in order to remain satisfiable $i(f) > i(g)$ must hold.
In a case where $f > g$ is not needed (but the TRS is KBO terminating)
the constraint must remain satisfiable. Thus $Y_{fg}$ or $Z_{fg}$ 
evaluate to one and because $i(g)$ is bound by $2^l - 1$ the constraint
does no harm. Summing up, the second constraint encodes a proper
order on the symbols in $\FF$. The third constraint forms an equivalence
relation on $\FF$ using the $Y_{fg}$ variables. Whenever $f \sim g$ is
demanded somehow in the encoding, then $X_{fg}$ and $Z_{fg}$ evaluate
to false by the first constraint. Satisfiability of the third constraint
implies $i(f) \geqslant i(g)$ but at the same time symmetry demands
that $Y_{gf}$ also evaluates to true which leads to $i(g) \geqslant i(f)$
and thus to $i(f) = i(g)$.

\begin{definition}
Let $\RR$ be a TRS. The pseudo-boolean encoding $\m{KBO}$-$\m{PBC}(\RR)$
is defined as the combination of $\m{ADM}$-$\m{PBC}(w,w_0)$,
$\m{PREC}$-$\m{PBC}(\FF)$, and
\[
\m{PBC}(l >_\kbo r); ~ \KBO_{l,r} = 1
\]
for all $l \to r \in \RR$.
\end{definition}

\begin{theorem}
\label{kbothm2}
A TRS $\RR$ is terminating whenever the PBCs
$\m{KBO}$-$\m{PBC}(\RR)$ are satisfiable.
\qed
\end{theorem}

Again the reverse does not hold (Example~\ref{unbounded bits}).

\section{Experimental Results}
\label{EXP:main}

We implemented our encodings on top of \TTT~\cite{TTT}.
MiniSat and MiniSat+ \cite{ES03,ES06} were used to check satisfiability of
the SAT and PBC based encodings.
Below we compare our implementations of KBO, \tsf{sat} and 
\tsf{pbc}, with the
ones of \TTT\ and \APROVE~\cite{APROVE}. \TTT\ admits only
strict precedences, \APROVE\ also quasi-precedences.
Both implement the polynomial time algorithm of Korovin and
Voronkov~\cite{KV03} together with techniques of
Dick \textsl{et al.}~\cite{DKM90}.

We used the 865 TRSs which do not specify any strategy or theory and the
322 string rewrite systems (SRSs) in version 3.2 of the Termination
Problem Data Base \cite{TPDB}. All tests were performed on a server
equipped with an Intel\textregistered{} Xeon\texttrademark{} processor
running at a CPU rate of 2.40 GHz and 512 MB of system memory with
a timeout of 60 seconds.

\subsection{Results for TRSs}

As addressed in
Section~\ref{ENC:kbo_main} one has to fix the number $k$ of bits which is
used to represent natural numbers in binary representation. The actual
choice is specified as argument to \tsf{sat} (\tsf{pbc}). Note that a
rather small $k$ is sufficient to handle all systems from \cite{TPDB}
which makes Theorems~\ref{kbothm1} and \ref{kbothm2} powerful in practice.
The example below gives evidence that there does not exist a general
upper bound on $k$.

\begin{example}
\label{unbounded bits}
Consider the parametrized TRS consisting of the three rules
\begin{xalignat*}{5}
\m{f}(\m{g}(x,y)) &\to \m{g}(\m{f}(x),\m{f}(y)) &
\m{h}(x)   &\to \m{f}(\m{f}(x))&
\m{i}(x) \to \m{h}^n(x) 
\end{xalignat*}
with $n = 2^k$. Since the first rule duplicates the function symbol
$\m{f}$ we must assign weight zero to it. The admissibility condition for
the weight function demands that $\m{f}$ is a maximal element in the
precedence. The second rule excludes the case $\m{h} \sim \m{f}$ and
demands that
the weight of $\m{h}$ is strictly larger than zero. It follows that the
minimum weight of $\m{h}^n(x)$ is $n+1 = 2^k+1$, which at the same time
is the minimum weight of $\m{i}(x)$. Thus $w(\m{i})$ is at least
$2^k$ which requires $k+1$ bits.
\end{example}

\begin{table}[tb]
\begin{tabular}{@{}lccc@{\quad}ccc@{}}
& \multicolumn{3}{c@{\qquad}}{strict precedence}
& \multicolumn{3}{c@{}}{quasi-precedence} \\
method(\#bits)
& total time & \#successes & \#timeouts
& total time & \#successes & \#timeouts \\
\hline
\tsf{sat/pbc}(2)
& 19.2/16.4 & 72/76 & 0/0
& 20.9/16.8 & 73/77 & 0/0 \\
\tsf{sat/pbc}(3)
& 20.2/16.3 & 77/77 & 0/0
& 21.9/16.9 & 78/78 & 0/0 \\
\tsf{sat/pbc}(4)
& 21.9/16.1 & 78/78 & 0/0
& 22.8/17.0 & 79/79 & 0/0 \\
\tsf{sat/pbc}(10)
& 86.1/16.7 & 78/78 & 1/0
& 90.2/17.2 & 79/79 & 1/0 \\
\TTT
& 169.5     & 77    & 1 
\end{tabular}
\medskip
\caption{KBO for 865 TRSs.}
\label{KBO:trs}
\vspace{-3ex}
\end{table}
The left part of Table~\ref{KBO:trs}
summarizes\footnote{The experiments are
described in more detail at \\
\texttt{http://cl-informatik.uibk.ac.at/\~{}hzankl/kbo}.}
the results for strict precedences.
Since \APROVE\ produced seriously slower results than \TTT\ in the
TRS category, it is not
considered in Table~\ref{KBO:trs}. Interestingly, with $k = 4$ equally
many TRSs can be proved terminating as with $k = 10$. The TRS
\tsf{higher-order\_AProVE\_HO\_ReverseLastInit} needs weight eight for
the constant \tsf{init} and therefore can only be proved KBO
terminating with $k \geqslant 4$.

Concerning the optimizations in Section~\ref{OPT:log},
if we use the standard (exponential) transformation to CNF, the
total time required increases to 2681.06 seconds, the number of
successful termination proofs decreases to 69, and 45 timeouts occur
(for $k = 4$).
Furthermore, if we don't use a cache for adding weights and equal symbols
are not removed when the weights of left and right-hand sides of
rules are compared, the number of successful termination proofs
remains the same but the total time increases to 92.30 seconds and one
timeout occurs.

\TTT\ without timeout requires 4747.65 seconds and can prove
KBO termination of 78 TRSs. The lion's share is taken up
by \tsf{various\_21} with 4016.23 seconds for a positive result.
\tsf{sat}(4) needs only 0.10 seconds for this TRS and \tsf{pbc}(4)
even only 0.03 seconds.
Since \TTT\ employs the slightly stronger KBO definition
of \cite{KV03} it can prove one TRS (\tsf{various\_27}) terminating
which cannot be handled by \tsf{sat} and \tsf{pbc}. On the other hand
\TTT\ gives up on \tsf{HM\_t000} which specifies addition for natural 
numbers in decimal notation (using 104 rewrite rules).
The problem is not the timeout but at some point the algorithm
detects that it will require too many resources. To prevent a likely
stack overflow from occurring, the computation is terminated and a
``don't know'' result is reported. (\APROVE\ behaves in a similar
fashion on this TRS.) Also for our approaches this system is the most
challenging one with 0.54 (\tsf{sat}(4)) and 0.11 (\tsf{pbc}(4)) seconds.


%

As can be seen from the right part of Table~\ref{KBO:trs}, by admitting
quasi-precedences one additional TRS (\tsf{SK\_90.2.42},
Example~\ref{example}) can be proved KBO terminating.
Surprisingly, \APROVE\ 1.2 cannot prove (quasi) KBO termination 
of this system, for unknown reasons. 

\subsection{Results for SRSs}

\begin{table}[tb]
\begin{tabular}{@{}lccc@{\quad}ccc@{}}
& \multicolumn{3}{c@{\qquad}}{strict precedence}
& \multicolumn{3}{c@{}}{quasi-precedence} \\
method(\#bits)
& total time & \#successes & \#timeouts
& total time & \#successes & \#timeouts \\
\hline
\tsf{sat/pbc}(2)
& \phantom{1}9.1/5.9 & \phantom{1}8/19 & 0/0
& 13.9/6.2 & \phantom{1}8/19 & 0/0 \\
\tsf{sat/pbc}(3)
& 12.1/5.9 & 17/24 & 0/0
& 16.9/6.4 & 17/24 & 0/0 \\
\tsf{sat/pbc}(4)
& 15.1/6.0 & 24/30 & 0/0
& 20.0/6.5 & 24/30 & 0/0 \\
\tsf{sat/pbc}(6)
& 15.8/6.1 & 31/33 & 0/0
& 27.4/6.7 & 31/33 & 0/0 \\
\tsf{sat/pbc}(7)
& 17.0/6.1 & 33/33 & 0/0
& 31.2/6.7 & 33/33 & 0/0 \\
\tsf{sat/pbc}(10)
& 21.6/6.3 & 33/33 & 0/0
& 98.8/6.9 & 32/33 & 1/0  \\
\TTT
& 72.4     & 29    & 1   
\end{tabular}
\medskip
\caption{KBO for 322 SRSs.}
\label{KBO:comp_srs}
\vspace{-3ex}
\end{table}
For SRSs we have similar results, as can be inferred from
Table~\ref{KBO:comp_srs}. The main
difference is the larger number of bits needed for the
propositional addition of the weights. The maximum number of SRSs is
proved KBO terminating with $k \geqslant 7$ in case of \tsf{sat} and
$k \geqslant 6$ for \tsf{pbc}. The reason is that in the first 
implementation the number of bits does not increase for intermediate 
sums when adding the weights. Generally speaking \TTT\
performs better on SRSs than on TRSs concerning KBO because it can 
handle all systems within 546.43 seconds. The instance which consumes 
the most time is \tsf{Zantema\_z112} with 449.01 seconds for a positive
answer; \tsf{sat}(7) needs just 0.11 and \tsf{pbc}(7) 0.03 seconds.
With a timeout of 60
seconds \TTT\ proves KBO termination of 29 SRSs, without any timeout
one more. Our implementations both prove KBO termination of 33 SRSs.
The three SRSs that make up the difference (\tsf{Trafo\_dup11},
\tsf{Zantema\_z069}, \tsf{Zantema\_z070}) derive from algebra
(polyhedral groups). \TTT\ and \APROVE\ give up on these SRSs for the 
same reasons as mentioned in the preceding subsection for \tsf{HM\_t000}.

Admitting quasi-precedences does not allow to prove KBO termination
of more SRSs. On the contrary, a timeout occurs when using
\tsf{sat}(10) on \tsf{Trafo\_dup11} whereas \tsf{pbc}(10) easily 
handles the system.

\section{Assessment}
\label{COM:main}

In this section we compare the two approaches presented in this paper.
Let us start with the most important measurements: power and run time.
Here \tsf{pbc} is the clear winner. Not only is it faster on any
kind of precedence; it also scales much better for larger numbers of
bits used to represent the weights. Furthermore, the pseudo-boolean
approach is less implementation work since additions are performed by
the SAT solver and also the transformation to CNF is not necessary.
We note that the implementation of \tsf{pbc} is exactly as described in
the paper whereas \tsf{sat} integrates the optimizations described
in Section~\ref{OPT:main}. 

A further advantage of the pseudo-boolean approach is the option of a 
\emph{goal function} which should be minimized while preserving
satisfiability of the 
constraints. Although the usage of such a goal
function is not of computational interest it is useful for generating
easily human readable proofs. We experimented with functions minimizing
the weights for function symbols and reducing the comparisons in the
precedence. The former has the advantage that one obtains a KBO proof
with minimal weights
which is nicely illustrated on the SRS \tsf{Zantema\_z113} consisting
of the rules
\begin{xalignat*}{3}
\m{11} &\to \m{43} &
\m{33} &\to \m{56} &
\m{55} &\to \m{62} \\
\m{12} &\to \m{21} &
\m{22} &\to \m{111} &
\m{34} &\to \m{11} \\
\m{44} &\to \m{3}  &
\m{56} &\to \m{12} &
\m{66} &\to \m{21}.
\end{xalignat*}
\TTT\ and \APROVE\ produce the proof 
\begin{xalignat*}{5}
w(1) &= 32471712256 & w(2) &= 48725750528 & w(3) &= 43247130624 \\
w(4) &= 21696293888 & w(5) &= 44731872512 & w(6) &= 40598731520 \\
&& 3 > 1 > 2 && 1 > 4
\end{xalignat*}
whereas \tsf{pbc}(6) produces
\begin{xalignat*}{5}
w(1) &= 31 & w(2) &= 47 & w(3) &= 41 \\
w(4) &= 21 & w(5) &= 43 & w(6) &= 39 \\
3 > {}& 1 > 2 & 3 > 5 &{} > 6 > 2 && 1 > 4.
\end{xalignat*}
Regarding the goal function dealing with the minimization of
comparisons in the precedence we detected that using two 
(three, four five, ten) bits to encode weights of function symbols 
39 (45, 46, 47, 47) TRSs can be proved terminating in 
16.7 (16.8, 17.0, 16.7, 16.9) seconds with empty precedence.

While running the experiments, \tsf{sat} and \tsf{pbc} produced different
answers for the SRS \tsf{Zantema\_z13}; \tsf{pbc} claimed KBO termination
whereas \tsf{sat} answered 
``don't know''. Chasing that discrepancy revealed a bug \cite{MBUG} in
MiniSat+ 
(which has been corrected in the meantime).

An interesting (and probably computationally fast) extension will be the
integration of the pseudo-boolean encoding of KBO into a dependency pair
\cite{AG00} setting
\cite{CSLTG06,ZHM07}. Clearly the constraints will get more
involved but we expect that the generalization to non-linear constraints
in the input format for the PB track of the SAT 2007 competition will
ease the work considerably.


\section{Summary}
\label{SUM:main}

In this paper we presented two logic-based encodings of KBO---pure SAT
and PBC---which can be implemented more efficiently and with
considerably less effort than the methods described in \cite{DKM90,KV03}.
Especially the PBC encoding gives rise to a very fast implementation
even without caring about possible optimizations in the encoding.

\bibliographystyle{plain}
\selectlanguage{english}

\end{document}